\documentclass[twoside]{article}
\usepackage{fleqn,espcrc2}
\usepackage{graphicx}

\newcommand{\AmS}{{\protect\the\textfont2
A\kern-.1667em\lower.5ex\hbox{M}\kern-.125emS}}	
\def\beq{\begin{equation}}
\def\eeq{\end{equation}}
\def\bea{\begin{eqnarray}}
\def\eea{\end{eqnarray}}
\def\bq{\begin{quote}}
\def\eq{\end{quote}}

\def\nnb{\nonumber}
\def\ga{\left(}
\def\dr{\right)}

\def\lb{\lbrack}

\def\rar{\rightarrow}

\def\nnb{\nonumber}
\def\la{\langle}
\def\ra{\rangle}
\def\nin{\noindent}
\def\ba{\vspace*{-0.2cm}\begin{array}}
\def\ea{\end{array}\vspace*{-0.2cm}}

\def\b{$\bullet~$}
\def\als{\alpha_s}

\def\gg2{ \la\alpha_s G^2 \ra}
\def\gg3{g^3f_{abc}\la G^aG^bG^c \ra}
\def\ggg4{\la\als^2G^4\ra}

\def\beq{\begin{equation}}
\def\enq{\end{equation}}
\def\beqa{\begin{eqnarray}}
\def\enqa{\end{eqnarray}}

\def\qq{\lag\bar{q}q\rag}
\def\sss{\lag\bar{s}s\rag}
\def\mix{\lag\bar{q}Gq\rag}
\def\mixs{\lag\bar{s}Gs\rag}

\def\G3{\lag g^3G^3\rag}

\def\al{\alpha}

\def\nn{\nonumber}
\def\lb{\label}

\newcommand{\rag}{\rangle}
\newcommand{\lag}{\langle}

\title
{\bf{\boldmath
{\Large $ SU(3)$  mass-splittings of heavy-baryons in QCD} }}

\author{
R.M. Albuquerque\,\thanks{{\it E-mail addresses:}, rma@if.usp.br (R.M. 
Albuquerque), snarison@yahoo.fr (S. Narison), mnielsen@if.usp.br (M. 
Nielsen).}\,\address {\footnotesize Instituto de F\'\i sica, 
Universidade de S\~ao Paulo, C.P. 66318, 05389-970 S\~ao Paulo, SP, 
Brazil.},
S. Narison\,\thanks{Corresponding author.}\,\address {\footnotesize 
Laboratoire
de Physique Th\'eorique et Astroparticules, CNRS-IN2P3 \& Universit\'e
de Montpellier II, Case 070, Place Eug\`ene
Bataillon, 34095 - Montpellier Cedex 05,
France. },
M. Nielsen\,$^a$,
}

\begin{document}

\pagestyle{myheadings}
\markright{ }
\begin{abstract}
\noindent
We extract directly (for the first time) the  charmed $(C=1)$ and bottom $(B=-1)$ heavy-baryons (spin
$1/2$ and $3/2$) mass-splittings due to $SU(3)$ breaking  using double 
ratios of QCD spectral sum rules (QSSR) in full QCD, which are less 
sensitive to the exact value and definition of the heavy quark mass, to the perturbative radiative
corrections
and to the QCD continuum contributions than the simple ratios 
commonly used for determining the 
heavy baryon masses. 
Noticing that most of the mass-splittings are mainly controlled by the ratio $\kappa\equiv\la \bar ss\ra/\la \bar 
dd\ra$ of the  condensate, we extract this ratio, 
 by allowing 1$\sigma$ deviation from  the 
observed masses of the $\Xi_{c,b}$ and of the $\Omega_c$. We obtain: $\kappa=0.74(3)$, which improves the existing estimates: $\kappa=0.70(10)$ from
light hadrons. Using this value, we deduce $M_{\Omega_b}=6078.5(27.4)$ MeV which agrees with the recent CDF data but disagrees by 2.4$\sigma$ with the one from D0. Predictions of the $\Xi'_Q$ and of the spectra of spin 3/2 baryons containing one or two strange quark are given in Table \ref{tab:mass}. 
Predictions of the hyperfine splittings $\Omega^*_Q-
\Omega_Q$ and  $\Xi^*_Q-\Xi_Q$ are also given in Table \ref{tab:hyperfine}. Starting for a general choice of the interpolating currents for the spin 1/2 baryons, our analysis favours 
 the  optimal value of the mixing angle $b\simeq (-1/5\sim 0)$  found from light and non-strange 
heavy baryons. 
 \end{abstract}
\maketitle
\vspace*{-1.5cm}
\section{Introduction}
\vspace*{-0.25cm}
 \nin
QSSR \cite{SNB,BB} \`a la SVZ \cite{SVZ} has been used earlier in full 
QCD \cite{BAGAN,BAGAN2,BAGAN3} and in HQET \cite{BAGAN4} for understanding 
heavy baryons [charmed ($cqq$), bottom ($bqq$),  double charm $(ccq)$, double bottom $(bbq)$ and $(bcq)$] 
masses. Recent observations at Tevatron of 
families of $b$-baryons \cite{TEV,PDG}  and of the $\Omega_c^*$ baryon by 
Babar and Belle \cite{BABAR} have stimulated different recent theoretical 
activities for understanding their nature
\cite{RICHARD,MARQUES,HUSSAIN,MARINA,SR,KORNER,EBERT,LATT,JENKINS}. 
QSSR results are in quite good agreement with recent experimental findings 
but with relatively large uncertainties. The inaccuracy of these results 
is mainly due 
to the value of the heavy quark mass and of its ambiguous definition when 
working to lowest order (LO) in the radiative $\alpha_s$ corrections in full QCD and 
HQET\,$^1$\footnotetext[1]{Radiative corrections to the heavy baryon correlators have been 
evaluated in \cite{KORNER} but for a particular choice of 
the interpolating currents.}, 
where the heavy quark mass is the main driving term in the QCD expression 
of the baryon two-point correlator used in the QSSR analysis. Another 
source of uncertainty is the effect of the QCD continuum which 
parametrizes the higher baryon masses contributions to the spectral 
function and the {\it ad hoc} choices of interpolating baryon currents used 
in different literatures. In this paper, we shall concentrate on the 
analysis of the heavy baryons mass-splittings due to $SU(3)$ breaking 
using double ratios (DR) of QCD spectral sum rules (QSSR), which are less 
sensitive to the exact 
value and definition of the heavy quark mass and to the QCD continuum 
contributions than the simple ratios used in the literature to determine 
the absolute value of heavy baryon masses. \\
In this letter, we extend the previous analysis in \cite{BAGAN,BAGAN2} by 
including the new $SU(3)$ breaking terms: $m_s$ and the ratio of the  condensate
$\kappa\equiv\la \bar ss\ra/\la \bar 
dd\ra$.
\\
\b {\bf For the spin 1/2 baryons},  and following Ref. \cite{BAGAN}, 
we work with the lowest dimension general currents:
\bea
\eta_{\Xi_Q}&=&\epsilon_{abc}\left[(q_a^TC\gamma_5s_b)+b(q_a^TCs_b)
\gamma_5\right]
Q_c~, \nnb\\
\eta_{\Lambda_Q}&=&\eta_{\Xi_Q}~~~~(s\rar q)~,\nnb\\
\eta_{\Omega_Q}&=&\epsilon_{abc}\left[(s_a^TC\gamma_5Q_b)
+b(s_a^TCQ_b)
\gamma_5\right]
s_c~, \nnb\\
\eta_{\Sigma_Q}&=&\eta_{\Omega_Q}~~~~(s\rar q)~,\nnb\\
\eta_{\Xi^\prime_Q}&=&{1\over\sqrt{2}}\epsilon_{abc}\Big{[}(s_a^TC
\gamma_5Q_b)q_c +
(q_a^TC\gamma_5Q_b)s_c\nnb\\
&&+b\left((s_a^TCQ_b)\gamma_5q_c + (q_a^TCQ_b)\gamma_5
s_c\right)\Big{]}~,
\label{cur1}
\eea
where we use standard notations; $b$ is {\it a priori} an arbitrary mixing parameter. Its value has 
been found to be:
\beq
b=-1/5~,
\label{eq:mixing1}
\eeq
in the case of light baryons \cite{JAMI2} and in the range 
\cite{BAGAN,BAGAN2,BAGAN3}:
\beq
-0.5\leq b\leq 0.5~,
\label{eq:mixing2}
\eeq
for non-strange heavy baryons, which do not favour the Ioffe choice $b=-1$ \cite{HEID}.
The corresponding two-point correlator reads:
\bea
S(q)&=&i\int d^4x~ e^{iqx}~ \la 0\vert {\cal T} \overline{\eta}_{Q}(x)
\eta_{Q}(0)\vert 0\ra\nnb\\
&\equiv& \hat q F_1  +F_2~,
\eea
where $F_1$ and $F_2$ are two invariant functions. \\
\b {\bf For the spin 3/2 baryons}, we  follow  Ref. \cite{BAGAN2} 
and work with the interpolating currents:
\bea
\eta^\mu_{\Xi^*_Q}&=&
\sqrt{2\over3}\big{[}
(q^TC\gamma_\mu Q)s+(s^TC\gamma_\mu Q)q+(q^TC\gamma_\mu s)Q\big{]}\nnb\\
\eta^\mu_{\Omega^*_Q}&=&{1\over\sqrt{2}}\eta^\mu_{\Xi^*_Q} ~~~~(q\rar s)
\nnb\\
\eta^\mu_{\Sigma^*_Q}&=&{1\over\sqrt{2}}\eta^\mu_{\Xi^*_Q}~~~~(s\rar q)~,
\label{cur}
\eea
where an anti-symmetrization over colour indices is understood.
The normalization in Eq.~(\ref{cur}) is chosen in such a way that in all 
cases
one gets the same perturbative contribution.
The corresponding two-point correlator reads:
\bea
S^{\mu\nu}(q)&=&i\int d^4x~ e^{iqx}~ \la 0\vert {\cal T} \overline{\eta}^\mu_{Q}(x)\eta^\nu_{Q}(0)\vert 0\ra\nnb\\
&\equiv& g^{\mu\nu}\ga \hat q F_1  +F_2\dr+\dots~ 
\eea
In the following, the contribution of the heavy quark
condensate $m_Q\la \bar QQ\ra$ will not appear\,\cite{BAGAN2} as it is cancelled by a part of the gluon condensate contribution due
to the heavy quark relation $m_Q\la \bar QQ\ra +(1/12\pi)\la\alpha_s G^2\ra\simeq 0$\,\cite{SVZ,SNB}. Again due to this relation,
the one due to $m_s\la \bar QQ\ra$  is numerically negligible because of the extra $(1/12\pi)$ loop and $1/M_Q$ factors compared to the one due to $m_s\la\bar qq\ra$ and due to the four-quark condensate contributions. The same loop factor also numerically suppresses the contributions of $\la\alpha_s G^2\ra$ and $m_s\la\alpha_s G^2\ra$ compared to the other ones. These negligible $SU(3)$ breaking contributions will not be considered in the following.  
\section{The spin 1/2 two-point correlator in QCD}
\label{sec:qcd1}
\vspace*{-0.35cm}
 \nin
 \subsection*{\b The \boldmath $\Lambda_Q(Qqq)$ and $\Xi_Q(Qsq)$ baryons}
 \nin
 The expression for $\Lambda_Q$ has been (first) obtained in the chiral 
limit $m_q=0$ in \cite{BAGAN2}, and the one of $\Xi_Q$ including $SU(3)$ 
breaking in \cite{MARINA}. One can notice that due to the expression of
 the current the $m_s$ corrections vanish to leading order in $\alpha_s$ 
for the perturbative term, while the $D=6$ condensates
for the $SU(2)$ case of \cite{BAGAN2} needs the following replacement  in 
the $SU(3)$ case:
\beq
\rho\la\bar qq\ra^2\rar \rho\la \bar qq\ra\la\bar ss\ra~,
\eeq
where $\rho=2\sim 3$ indicates the violation of the four-quark vacuum 
saturation \cite{LNT,SNB,SNTAU}.
 The additional $SU(3)$ breaking corrections for the $\Xi_Q$ are 
\cite{MARINA}:\\ 
{\boldmath  {\rm -~$ F_1: $}}
{\small
\bea
 {\rm Im}F_1^{m_s}\vert_{\bar ss}&=&-{m_s\over2^4\pi}(1-x^2) \Bigg{[} 
(1-b^2)\la \bar qq\ra-
\nnb\\ &&{(1+b^2)\over 2}\la \bar ss\ra\Bigg{]},
\nn\\
F_1^{m_s}\vert_{mix}&=&{m_s\over2^5\pi^2}{1\over m_Q^2-q^2}
\Bigg{\{} \mixs\ {(1+b^2)\over 6}+ 
\nnb\\
&&\la\bar  qGq\ra(1-b^2)\Bigg{\}}, 
\label{eq:f1chi}
\eea
}
{\boldmath  {\rm -~$ F_2: $}}
{\small
\bea
{\rm Im}F_2^{m_s}\vert_{\bar ss}&=&-{m_sm_Q\over2^3\pi}(1-x) \Bigg{[} 
(1+b^2)\la \bar qq\ra-
\nnb\\ &&{(1-b^2)\over 2}\la \bar ss\ra\Bigg{]},
\nn\\
F_2^{m_s}\vert_{mix}&=&{m_sm_Q\over2^5\pi^2}{1\over m_Q^2-q^2}\Bigg{\{} 
\mixs\ {(1-b^2)\over 6}+ \nnb\\
&&\la\bar  qGq\ra(1+b^2)\Bigg{\}},
\label{eq:f2chi}
\eea
}
where $x\equiv m_Q^2/s$ and $\mixs\equiv g\la \bar s\sigma_{\mu\nu}
\lambda_a/2G^{\mu\nu}_a s\ra$ . 
%
 \subsection*{\b The \boldmath $\Sigma_Q(Qqq)$ and $\Omega_Q(Qss)$ baryons}
 \nin
  The expression for $\Sigma_Q$ has been (first) obtained  in 
\cite{BAGAN}. The additionnal $SU(3)$ breaking terms for the $\Omega_Q$ 
are:\\
 {\boldmath  {\rm -~$ F_1: $}}
 {\small
 \bea
 {\rm Im} F_1^{m_s}\vert_{pert}&=&{3m_sm_Q^3\over2^8\pi^3}(1-b^2)\times
\nnb\\
 && \Bigg{[}{2\over x}+3-6x+x^2
+6\ln{x} \Bigg{]},
\nn\\
{\rm Im}F_1^{m_s}\vert_{\bar ss}&=&{3m_s\sss\over2^6\pi}(1+b)^2\left(1-x^2
\right),
\nn\\
F_1^{m_s}\vert_{mix}&=&-{m_s\mixs\over2^73\pi^2}
 \Bigg{[}{1\over m_Q^2-q^2}
(7+22b+7b^2)
\nnb \\  &&
-6(1+b)^2
 \int_0^1{d\al(1-\al)\over m_Q^2-(1-\al)q^2}
\Bigg{]}~,
\nn\\
F_1^{m_s}\vert_{D=6}&=&-{m_sm_Q\rho\la\bar ss\ra^2(1-b^2)\over 8
(m_Q^2-q^2)^2}~.
\eea
}
{\boldmath  {\rm -~$ F_2: $}}
{\small
\bea
{\rm Im}F_2^{m_s}\vert_{pert}&=&{3m_sm_Q^4\over2^8\pi^3}(1-b^2)
\times\nnb\\ &&
\left({1\over x^2}-{6\over 
x}+3+2x-6\ln{x}\right),
\nn\\
{\rm Im}F_2^{m_s}\vert_{\bar ss}&=&-{3m_sm_Q\sss\over2^5\pi}(3+2b+3b^2)
\left(1-x\right),
\nn\\
F_2^{m_s}\vert_{mix}&=&{m_sm_Q\mixs\over2^73\pi^2}\times\nnb\\
&&\Bigg{[}{1\over m_Q^2-q^2}
(25+22b+25b^2)\nn\\
&&-3(5+6b+5b^2)\times\nnb\\
&&\int_0^1{d\al\over m_Q^2-(1-\al)q^2}
\Bigg{]}~,
\nn\\
F_2^{m_s}\vert_{D=6}&=&-{m_s\rho\la\bar ss\ra^2(1-b^2)\over 8(m_Q^2-q^2)}
\Bigg{[} 1+{m_Q^2\over 
m_Q^2-q^2}\Bigg{]}~,
\nn\\
\eea 
}
 \subsection*{\b The \boldmath $\Sigma_Q(Qqq)$ and $\Xi'_Q(Qsq)$ baryons}
 \nin
 The expression for the $\Xi'_Q$ tends to the one of the $\Sigma_Q$ in the chiral
 limit $m_{q,s}\to 0$ and is very similar with one of the $\Omega_Q$. The SU(3) breaking
 corrections read:\\
{\boldmath $F_1~:$}
\beqa\label{estq}
%
%
\mbox{Im }F_1^{m_s}|_{pert}&=&{3m_sm_Q^3\over2^9\pi^3}(1-b^2)\times\nnb\\
&&\Bigg{[}{2\over
x}+3-6x+x^2+6\ln{x}\Bigg{]},
\nn\\
\mbox{Im }F_1|_{\bar ss}&=&-{3m_Q(\sss+\qq)\over2^6\pi}(1-b^2)(1-x)^2,
\nn\\
\mbox{Im }F_1^{m_s}|_{\bar ss}&=&{m_s\over2^7\pi}\left(1-x^2\right)
\Bigg{[}-2(1-
b)^2\qq\nnb\\
&&+(5+2b+5b^2)\sss\Bigg{]},
\nn\\
%
%
\mbox{Im }F_1|_{mix}&=&{(\mixs+\mix)\over m_Q2^8\pi}(1-b^2)\left(13x^2-
6x\right),
\nn\\
F_1^{m_s}|_{mix}&=&-{m_s\over2^83\pi^2}\Bigg{[}{1\over m_Q^2-q^2}\Big{[}(13+10b
+13b^2)\times\nnb\\
&&\mixs-6(1-b)^2\mix\Big{]}+\nnb\\
&& \Big[3(1-b)^2\mix\nnb\\
&&-3(3+2b+3b^2)\mixs\Big]\times\nnb\\
&& \int_0^1d\alpha {1-\alpha\over
m_Q^2-(1-\alpha)q^2}\Bigg],\nn\\
F_1|_{D=6}&=&{\rho\sss\qq\over24}(1-b)^2{1\over m_Q^2-q^2},\nn\\
F_1^{m_s}|_{D=6}&=&-{m_sm_Q\rho\sss\qq\over16}{1-b^2\over (m_Q^2-q^2)^2}
\enqa

\noindent

{\boldmath$F_2~:$}
\beqa\label{estm}
%
%
\mbox{Im }F_2^{m_s}|_{pert}&=&{3m_sm_Q^4\over2^9\pi^3}
(1-b^2)\times\nnb\\
&&\Bigg{[}{1\over x^2}-{6\over x}+3+2x-6\ln{x}\Bigg{]},
\nn\\
\mbox{Im }F_2|_{\bar ss}&=&-{3m_Q^2(\sss+\qq)\over2^6\pi}\times\nnb\\
&&(1-b^2)x\left(1-{1\over
x}\right)^2,
\nn\\
\mbox{Im }F_2^{m_s}|_{\bar ss}&=&{m_sm_Q\over2^6\pi}\left(1-x\right)\Big{[}(1-b)^2
\sss-\nnb\\
&& 2(5+2b+5b^2)\qq\Big{]},
\nn\\
%
%
\mbox{Im }F_2|_{mix}&=&{(\mixs+\mix)\over 2^8\pi}(1-b^2)\left(6+x\right),
\nn\\
F_2^{m_s}|_{mix}&=&-{m_sm_Q\over 2^83\pi^2}\Bigg[{1\over m_Q^2-q^2}\Big[
5(1-b)^2\mixs\nnb\\
&&-6(5+2b+5b^2)\mix\Big{]}+\nnb\\
&&\Big[-3(1-b)^2\mixs+\nnb\\
&&6(3+2b+3b^2)\times\nnb\\
&&\mix\Big]\int_0^1{d\alpha\over
m_Q^2-(1-\alpha)q^2}\Bigg],\nn\\
F_2|_{D=6}&=&{m_Q\rho\sss\qq\over24}(5+2b+5b^2){1\over m_Q^2-q^2},\nn\\
F_2^{m_s}|_{D=6}&=&-{m_s\rho\sss\qq\over16}{1-b^2\over m_Q^2-q^2}\times\nnb\\
&&\left[1
+{m_Q^2\over m_Q^2-q^2}\right].
\enqa
We have checked the existing results in \cite{BAGAN} obtained in the 
chiral limit and all our previous results agree with these ones. 
%

\vspace*{-0.3cm}
\section{The spin 3/2 two-point correlator in QCD}
\label{sec:qcd2}
\vspace*{-0.25cm}
 \nin
 The QCD expression of the two-point correlator for the $\Sigma^*_Q(Qqq)$ 
has been (first) obtained in the chiral limit $m_{u,d}=0$,   to LO  in 
$\alpha_s$ and up to the contributions of the $D=6$ condensates in 
\cite{BAGAN}. In this letter, we extend the previous analysis by including 
the new $SU(3)$ breaking $m_s$ correction terms and consider the $SU(3)$ 
breaking of the ratio of quark condensates $\sss\not=\qq$ like we did for the spin 1/2 case.
 %
 \subsection*{\b The  \boldmath $\Sigma^*_Q(Qqq)$ and $\Xi^*_Q(Qsq)$  
baryons}
 \nin
The additionnal terms and replacement due to $SU(3)$ breaking for the 
$\Xi^*_Q$ compared with
the one of the $\Sigma^*_Q(Qqq)$ in \cite{BAGAN2} are:\\
{\boldmath  {\rm -~$F_1: $}}
{\small
 \bea
 {\rm Im} F_1^{m_s}\vert_{pert}&=&{m_sm_Q^3\over48\pi^3}
 \Bigg{[}{2\over x}+3-6x+x^2
+6\ln{x} \Bigg{]}~,
\nn\\
{\rm Im}F_1\vert_{\bar ss}&=& -{m_Q\over 6\pi} \Big{[} \qq+\sss \Big{]}\ga 
1-x\dr^2~,
\nn\\
{\rm Im}F_1^{m_s}\vert_{\bar ss}&=&-{m_s\over 12\pi}\Bigg{[} 2(1-x^2)\qq-
(1-x^3)\sss \Bigg{]}~,
\nn\\
{\rm Im}F_1\vert_{mix}&=& {7M_0^2\over 3^2 2^3\pi} \Big{[} \qq+\sss \Big{]}
{x^2\over m_Q}~,
\nn\\
F_1^{m_s}\vert_{mix}&=&{m_sM_0^2\over144\pi^2}
\Bigg{[}{12\qq-9\sss \over m_Q^2-q^2}+
 \nnb\\ &&
2\int_0^1{d\al(1-\al)\over m_Q^2-(1-\al)q^2}\times \nnb\\
&&\Big{[} (1-3\al)\sss+\qq\Big{]}
\Bigg{]}~,
\nn\\
F_1\vert_{D=6}&=& {4\over 9}{\rho\sss\qq\over m_Q^2-q^2}~,
\nn\\
F_1^{m_s}\vert_{D=6}&=&-{2\over 9}m_Qm_s{\rho\sss\qq\over \ga m_Q^2-q^2
\dr^2}~.
\eea
}
{\boldmath  {\rm -~$F_2: $}}
{\small
\bea
{\rm Im}F_2^{m_s}\vert_{pert}&=&{m_sm_Q^4\over192\pi^3}
\Bigg{[}{3\over x^2}-{16\over 
x}+12+x^2-12\ln{x}\Bigg{]},
\nn\\
{\rm Im}F_2\vert_{\bar ss}&=& -{m_Q^2\over 18\pi} \Big{[} \qq+\sss \Big{]}
\ga {2\over x}-3+x^2\dr~,
\nn\\
{\rm Im}F_2^{m_s}\vert_{\bar ss}&=&-{m_sm_Q\over12\pi}(1-x)\Big{[} 6\qq-
(1+x)\sss\Big{]},
\nn\\
{\rm Im}F_2\vert_{mix}&=& {M_0^2\over 18\pi} \Big{[} \qq+\sss \Big{]}\ga 1+
{3\over 4}x^2\dr~,
\nn\\
F_2^{m_s}\vert_{mix}&=&{m_sm_QM_0^2\over72\pi^2}
\Bigg{[}3{(3\qq-\sss) \over m_Q^2-q^2}+
 \nnb\\ &&
\qq\int_0^1{d\al\over m_Q^2-(1-\al)q^2}\Bigg{]}~,
\nn\\
F_2\vert_{D=6}&=&{2\over 3}{m_Q\rho\sss\qq\over m_Q^2-q^2}~,
\nn\\
F_2^{m_s}\vert_{D=6}&=&-{2\over 9}{m_sm_Q^2\rho\sss\qq\over  (m_Q^2-q^2)^2}~,
\eea
}
where $x\equiv m_Q^2/s$ and $\mixs\equiv g\la \bar s\sigma_{\mu\nu}
\lambda_a/2G^{\mu\nu}_a s\ra\equiv M_0^2\sss$ . 
 \subsubsection*{\b The \boldmath $\Omega^*_Q(Qss)$ baryons}
 \nin
Compared with the expression of the $\Sigma^*_Q(Qqq)$ in\,\cite{BAGAN2}, 
the additionnal $SU(3)$ breaking terms for the $\Omega^*_Q$ are:\\
{\boldmath  {\rm -~$ F_1: $}}
{\small
 \bea
 {\rm Im} F_1^{m_s}\vert_{pert}&=&{m_sm_Q^3\over24\pi^3}
 \Bigg{[}{2\over x}+3-6x+x^2
+6\ln{x} \Bigg{]},
\nn\\
{\rm Im}F_1^{m_s}\vert_{\bar ss}&=&-{m_s\sss\over6\pi}\ga 1-2x^2+x^3\dr,
\nn\\
F_1^{m_s}\vert_{mix}&=&{m_sM_0^2\sss\over72\pi^2}
\Bigg{[}{3\over m_Q^2-q^2}+
 \nnb\\ &&
2\int_0^1{d\al(1-\al)(2-3\al)\over m_Q^2-(1-\al)q^2}
\Bigg{]}~,
\nn\\
F_1^{m_s}\vert_{D=6}&=&-{4\over 9}{m_Qm_s\rho\sss^2\over \ga m_Q^2-q^2
\dr^2}~.
\eea
}
{\boldmath  {\rm -~$F_2: $}}
{\small
\bea
{\rm Im}F_2^{m_s}\vert_{pert}&=&{m_sm_Q^4\over96\pi^3}\nnb\\
&&\Bigg{[}{3\over x^2}-{16\over 
x}+12+x^2-12\ln{x}\Bigg{]},
\nn\\
{\rm Im}F_2^{m_s}\vert_{\bar ss}&=&-{m_sm_Q\sss\over6\pi}\Bigg{[} 5-6x+x^2
\Bigg{]},
\nn\\
F_2^{m_s}\vert_{mix}&=&{m_sm_QM_0^2\sss\over36\pi^2} \Bigg{[}{6\over m_Q^2-
q^2}+
\nn\\
&&\int_0^1{d\al\over m_Q^2-(1-\al)q^2 }
\Bigg{]}~,
\nn\\
F_2^{m_s}\vert_{D=6}&=&-{4\over 9}{m_sm_Q^2\rho\sss^2\over(m_Q^2-q^2)^2}.
\eea
}
\\
We have checked the existing results in \cite{BAGAN2} obtained in the 
chiral and $SU(2)$ limits and agree with these ones. 
\vspace*{-0.3cm}
\section{Form of the sum  rules and QCD inputs}
\label{sec:qssr}
\vspace*{-0.25cm}
 \nin
We parametrize the spectral function using the standard duality ansatz: 
``one resonance"+ ``QCD continuum". The QCD continuum starts from a 
threshold $t_c$ and comes from the discontinuity of the QCD diagrams. 
Transferring its contribution to the QCD side of the sum rule, one obtains 
the finite energy Laplace/Borel sum rules:
\bea
&&|\lambda_{B^{(*)}_q}|^2M_{B^{(*)}_q}~e^{-{M_{B^{(*)}_q}}^2\tau}=
\int_{t_q}^{t_c}ds~
e^{-s\tau}~{1\over\pi}{\rm Im}F_{2}(s)~,\nnb\\
&&|\lambda_{B^*_q}|^2~e^{-{M_{B^{(*)}_q}}^2\tau}=\int_{t_q}^{t_c}ds~
e^{-s\tau}~{1\over\pi}{\rm Im}F_{1}(s)~,
\lb{srm} 
\eea
where $\tau\equiv 1/M^2$ is the sum rule variable and $\lambda_{B^{(*)}_q}$ and $M_{B^{(*)}_q}$ are the heavy baryon residue and mass. Notice that, within our choice of the interpolating currents, we may have a contamination due to the negative parity states contribution, which we can quantify from the ratio
of sum rules associated to $(M_QF_1-F_2)$ and $(M_QF_1+F_2)$ \cite{BAGAN4}. We shall check (a posteriori) that this ratio is negligible at the stability regions. On the other, the observed negative parity states are systematically heavier than the positive ones such that they can be legitimately included into the QCD continuum contribution. 
Consistently, we also take into account the $SU(3)$ breaking at the quark 
and continuum threshold:
 \bea
   \sqrt{t_q}\vert_{SU(3)}&\simeq&   \ga\sqrt{t_q}\vert_{SU(2)}\equiv m_Q
\dr+ \bar m_{q_1}+\bar m_{q_2}~,\nnb\\
   \sqrt{t_c}\vert_{SU(3)}&\simeq&   \ga\sqrt{t_c}\vert_{SU(2)}\equiv 
\sqrt{t_c}\dr + \bar m_{q_1}+\bar m_{q_2}~,
   \eea
where $q_{1,2}\equiv q$ or/and $s$ depending on the channel.  $\bar m_{q_i}
$  are the running light  quark masses. $m_Q$ is the heavy quark mass, 
which we shall take in the range covered by the running and on-shell mass 
(see Table \ref{tab:param}) because of its ambiguous definition  when 
working to lowest order of perturbative QCD. 
One can estimate the baryon masses from the following ratios:
\bea
&&{\cal R}^q_i={\int_{t_q}^{t_c}ds~s~
e^{-s\tau}~{\rm Im}F_{i}(s)\over \int_{t_q}^{t_c}ds~
e^{-s\tau}~{\rm Im}F_{i}(s)}~,~~~~~i=1,2~,\nnb\\
&&{\cal R}^q_{21}={\int_{t_q}^{t_c}ds~
e^{-s\tau}~{\rm Im}F_{2}(s)\over \int_{t_q}^{t_c}ds~
e^{-s\tau}~{\rm Im}F_{1}(s)}~,
\label{eq:ratio}
\eea
where at the $\tau$-stability point :
\beq
M_{B^{(*)}_q}\simeq \sqrt{{\cal R}^q_i}\simeq {\cal R}^q_{21}~.
\eeq
These quantities have been used in the literature for getting the baryon 
masses
and lead to a  typical uncertainty of 15-20\% \cite{BAGAN,BAGAN2,BAGAN3} 
\,$^2$\footnotetext[2]{More accurate results quoted in the recent QSSR literature 
\cite{MARINA,SR} do not take into account the uncertainties due to the 
heavy quark mass definitions and to the arbitrary choice of the baryonic 
interpolating currents.}. In order to circumvent these problems, we work 
with the double ratio of sum rules (DR)\cite{SNhl}:
\beq
r^{sd}_i\equiv \sqrt{{\cal R}^s_i\over {\cal R}^d_i}~,~~~~~r^{sd}_{21}
\equiv {{\cal R}^s_{21}\over {\cal R}^d_{21}}~.
\label{eq:2ratio}
\eeq
which take directly into account the $SU(3)$ breaking effects. These 
quantities are obviously less sensitive to the choice of the heavy quark 
masses, to the perturbative radiative corrections and to the value of the continuum threshold than the simple ratios 
${\cal R}_i$ and  ${\cal R}_{21}$\,$^2$\,\footnote{One may also 
work with the double ratio of moments ${\cal M}_n$ based on different 
derivatives  at $q^2=0$ \cite{SNhl}. However, in this case the OPE is 
expressed as an expansion in $1/m_Q$, which for a LO expression of the QCD 
correlator is more affected by the definition of the heavy quark mass to 
be used.}. Analogous DR quantities have been used successfully (for the 
first time) in \cite{SNhl} for studying the mass ratio of the $0^{++}/
0^{-+}$ and $1^{++}/1^{--}$ B-mesons, in \cite{SNFBS} for extracting  
$f_{B_s}/f_B$, in \cite{SNFORM} for estimating the $D\to K/D\to \pi$ 
semi-leptonic form factors and in \cite{SNe+e-} for extracting the strange 
quark mass from the $e^+e^-\rar I=1,0$ data. 
For the numerical analysis whe shall  introduce the RGI quantities $\hat
\mu$ and $\hat m_q$ \cite{FNR}:
\bea
\bar m_q(\tau)&=&{\hat m_q\over \ga -\log{ \sqrt{\tau}\Lambda}\dr^{2/{-
\beta_1}}}\nnb\\
{\la\bar qq\ra}(\tau)&=&-{\hat \mu_q^3 \ga-\log{ \sqrt{\tau}\Lambda}\dr^{2/
{-\beta_1}}}\nnb\\
{\la\bar qGq\ra}(\tau)&=&-{\hat \mu_q^3 \ga-\log{ \sqrt{\tau}\Lambda}
\dr^{1/{-3\beta_1}}}M_0^2~,
\eea
where $\beta_1=-(1/2)(11-2n/3)$ is the first coefficient of the $\beta$ 
function for $n$ flavours. We have used the quark mass and condensate 
anomalous dimensions reviewed in \cite{SNB}.  We shall use the QCD 
parameters in Table \ref{tab:param}. At the scale where we shall work, and 
using the paramaters in the table, we deduce:
\beq
\rho=2.1\pm 0.2~,
\eeq
which controls the deviation from the factorization of the four-quark 
condensates. 
We shall not include the $1/q^2$ term discussed in \cite{CNZ,ZAK},which is 
consistent with the LO approximation used here as the latter has been 
motivated for a phenomenological parametrization  of the larger order terms
 of the QCD series.

{\scriptsize
\begin{table}[hbt]
\setlength{\tabcolsep}{0.5pc}
 \caption{\scriptsize    QCD input parameters. For the heavy quark masses, 
we use the range spanned
 by the running $\overline{MS}$ mass $\overline{m}_Q(M_Q)$  and the 
on-shell mass from QSSR compiled in page
 602,603 of the book in \cite{SNB}. 
    }
    {\small
\begin{tabular}{lll}
&\\
\hline
Parameters&Values& Ref.    \\
\hline
$\Lambda$& $(353\pm 15)$ MeV &\cite{SNTAU,PDG}\\
$\hat m_d$&$(6.1\pm 0.5)$ MeV&\cite{SNmass,SNB,PDG}\\
$\hat m_s$&$(114.5\pm 20.8)$ MeV&\cite{SNmass,SNB,PDG}\\
$\hat \mu_d$&$(263\pm 7)$ MeV&\cite{SNmass,SNB}\\
$\kappa\equiv{\la\bar ss\ra}/{\la\bar dd\ra}$&$(0.7\pm 0.1)$&
\cite{SNmass,SNB,JAMI2} \\
$M_0^2$&$(0.8 \pm 0.1)$ GeV$^2$&\cite{JAMI2,HEID,SNhl}\\
$\la\alpha_s G^2\ra$& $(6.8\pm 1.3)\times 10^{-2}$ GeV$^4$&
\cite{SNTAU,LNT,SNI,fesr,YNDU,SNHeavy,BB}\\
$\rho\alpha_s\la \bar dd\ra^2$& $(4.5\pm 0.3)\times 10^{-4}$ GeV$^6$&
\cite{SNTAU,LNT}\\
$m_c$&$(1.18\sim1.47)$ GeV &\cite{SNB,SNmass,SNHmass,PDG}\\
$m_b$&$(4.18\sim4.72)$ GeV &\cite{SNB,SNmass,SNHmass,PDG}\\
\hline
\end{tabular}
\vspace*{-0.5cm}
}
\label{tab:param}
\end{table}
}
\nin
\section{\boldmath$\kappa\equiv\la \bar ss\ra/\la \bar 
dd\ra $ from  the spin 1/2 baryons }
\vspace*{-0.25cm}
  \nin
As a preliminary step of the analysis, we check the different results 
obtained in full QCD and in the chiral limit \cite{BAGAN,BAGAN2}:
\bea
&& M_{\Sigma_c}=(2.45\sim 2.94)~{\rm GeV}~,\nnb\\
&& M_{\Sigma_b}=(5.70\sim 6.62)~{\rm GeV}~,\nnb\\
&& M_{\Sigma_c}-M_{\Lambda_c}\leq207~{\rm MeV}~,\nnb\\
&& M_{\Sigma_b}-M_{\Lambda_b}\leq163~{\rm MeV}~,
\eea
which we confirm. However, we have not tried to improve these results for the absolute
values of the masses due 
to the ambiguity in the definition of the heavy quark mass
input mentioned earlier at LO, which induces large errors. 
 \subsection*{\b  \boldmath $\Xi_c(csq)/\Lambda_c(cqq)$}
 \nin
 \begin{figure}[hbt]
 \vspace*{-0.5cm}
\begin{center}
\includegraphics[width=6cm]{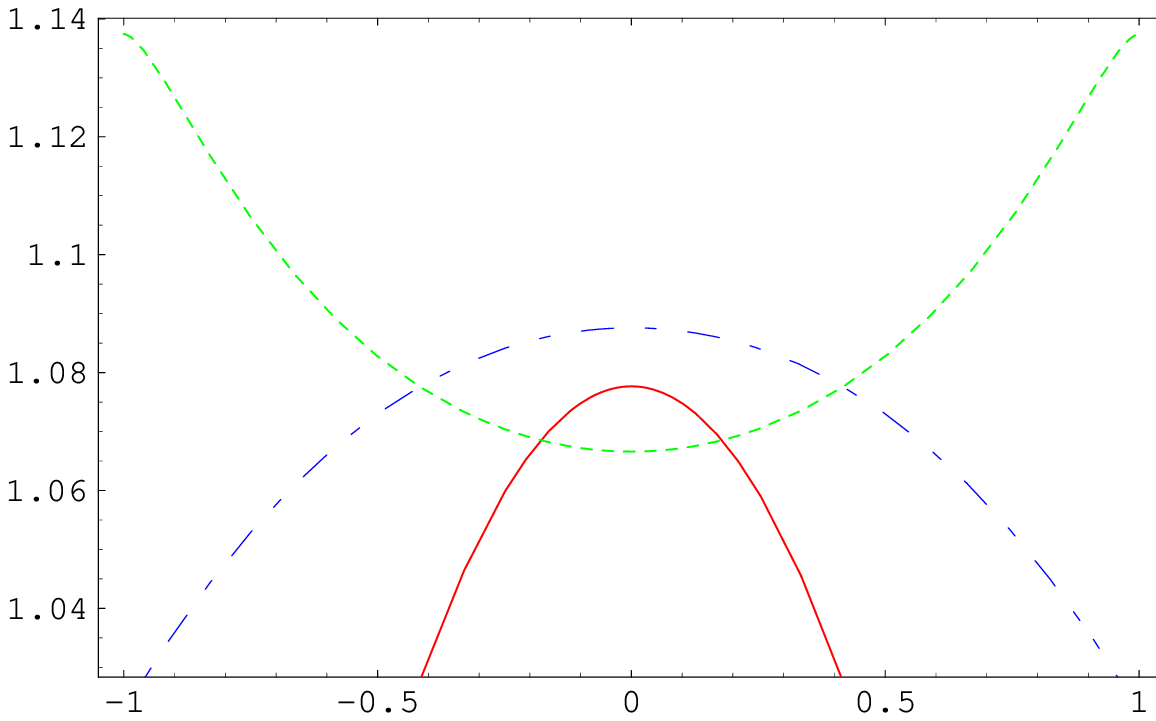}
\includegraphics[width=6cm]{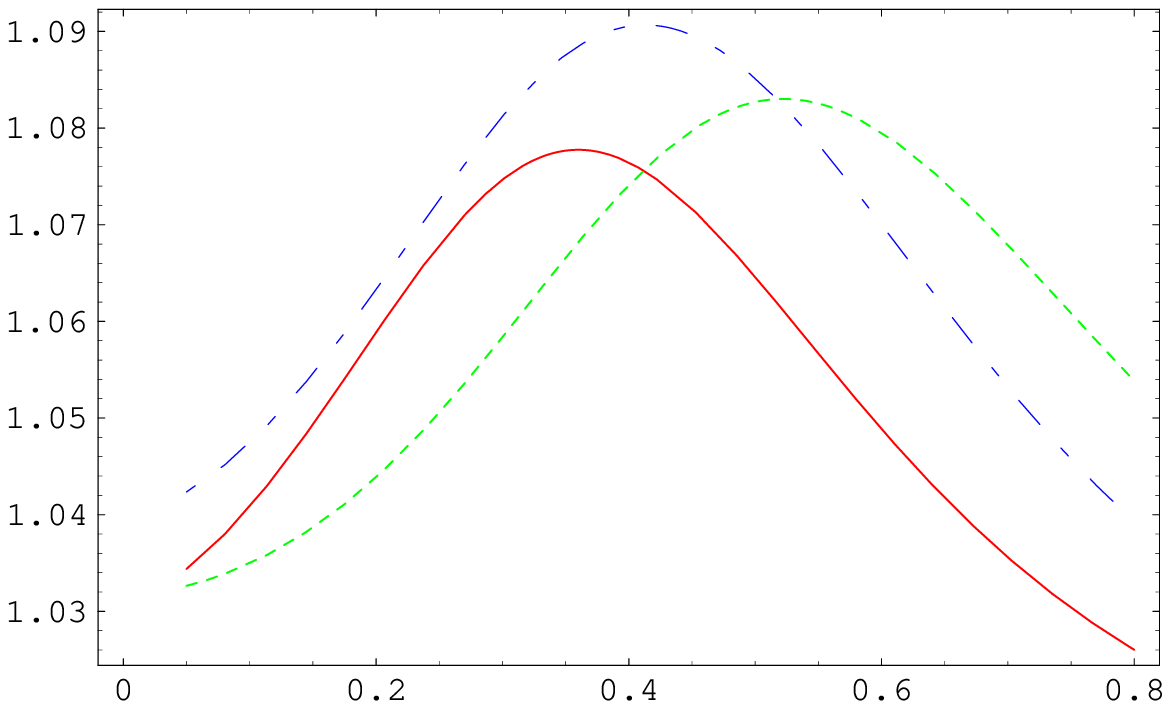}
\includegraphics[width=6cm]{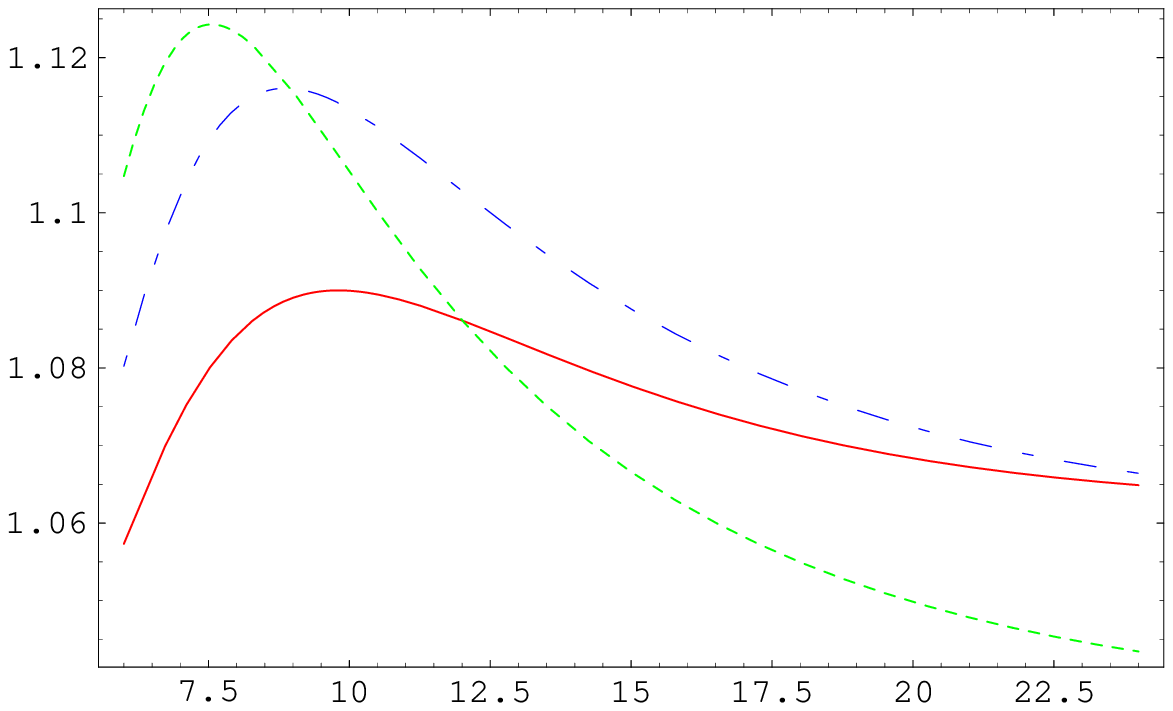}
\vspace*{-0.7cm}
\caption{\footnotesize {\boldmath $\Xi_c/\Lambda_c$}:  a) $b$-behaviour 
of the double ratio of sum rules (DR) given $\tau=0.35$ GeV$^{-2}$ and $t_c=
15$ GeV$^2$: $r_1^{ds}$ dashed-dotted (blue),  $r_2^{sd}$ dotted (green),  
$r_{21}^{sd}$ continuous (red); b) $\tau$-behaviour for 
$b=0$  and $t_c=15$ GeV$^2$; 
c) $t_c$-behaviour of the DR 
given $b=0$ and $\tau=0.35$ GeV$^{-2}$; We have used $\kappa=0.7$.   }
\label{fig:chic}
\end{center}
\vspace*{-1.cm}
\end{figure}
\nin
 {\it -- Optimal choice of the currents:} we start from the general choice of interpolating
 currents given in Eq. (\ref{cur1}). We study the $b$-behaviour of the predictions in 
  Fig. \ref{fig:chic}a), by 
fixing $\tau$=0.35 GeV$^{-2}$ and $t_c$= 15 GeV$^2$. The result  presents $b$-stability
around $b=0$, which is in the range given in 
Eq. (\ref{eq:mixing2}) obtained from light baryons and heavy non-strange baryons. 
We consider this value as the optimal choice of the interpolating currents.
However, this generous range does not favour the 
{\it ad hoc} choice around 1 used in the existing literature 
\cite{MARINA,SR}. 
Therefore, in this channel, we shall work with:
\beq
b\simeq 0~.
\label{eq:b}
\eeq
  {\it --  $\tau$ stabilities:}  we show in Fig. \ref{fig:chic}b) the $\tau$ 
behaviour of the different DR at fixed $tc=15$ GeV$^{-2}$ and $b=0$. \\
{\it -- $t_c$ stabilities and choice of the sum rules:} 
fixing $b=0$ from the previous analysis, we study in Fig. \ref{fig:chic}c)
the $t_c$-behaviour of predictions. Among the three sum rules, we retain
 $r_{21}^{sd}$ which is the most stable in $t_c$ and then less affected
by the higher state contributions. \\
 {\it -- Results:} 
From this $r_{21}^{sd}$ sum rule, we can deduce the DR:
\bea
r_{\Xi_c}^{sd}= 1.080(10)(2)(6)(2)(1.5)~,
\label{eq:rchic}
\eea
where the value $\kappa=0.7$ has been taken.
We have considered the mean value of $r_{21}^{sd}$ from $t_c=10$ to 20 GeV$^2$.  
The errors are due respectively to the values of $t_c$, $\tau=(0.35\pm 0.05)$ 
GeV$^{-2}$, $m_c$, $m_s$ and the factorization of the four-quark condensate $\rho$.
The errors due to $b$ and some other SU(3) symmetric QCD parameters are negligible.
The large error due to $\kappa$ compiled in Table \ref{tab:param} is not included in Eqs. (\ref{eq:rchic}).  
Using as input the data\,\cite{PDG}: 
\bea
M_{\Lambda_c}^{exp}=(2286.46\pm 0.14)~{\rm MeV}~,
\eea
and adding the different errors quadratically, one can deduce:
\bea
M_{\Xi_c}=(2469.4\pm 26.6)~{\rm MeV}~,
\eea
\label{eq:chic}
which agrees nicely with the data \cite{PDG}:
\bea
M_{\Xi_c}^{exp}=(2467.9\pm 0.4)~{\rm MeV}~.
\eea
For improving the existing value of $\kappa$ ,
we allow a $1\sigma$ deviation of the DR prediction from the experimental value. 
In this way, we deduce:
\bea
\kappa= 0.700(50)~. 
\label{eq:kchic}
\eea

 \subsection*{\b  \boldmath $\Xi_b(bsq)/\Lambda_b(bqq)$}
 \nin
We repeat the previous analysis in the case of the $b$-quark. The analysis 
of the ratio of sum rules  
shows similar curves than for the charm case except the obvious change of 
scale.  Using  ${r}_{21}^{sd}$, we illustrate the analysis using $\kappa=0.74$.
In this way, we obtain:
  \beq
 r_{\Xi_b}^{sd}=1.030(2.5)(0.5)(1.5)(0.5)(0.5)~,
 \eeq
where the errors come from $t_c$ from 45 to 80 GeV$^2$, $\tau=(0.18\pm 0.05)$ 
GeV$^{-2}$, $m_b$, $m_s$ and the factorization of the four-quark condensate $\rho$.
From this result, and using:
\bea
M_{\Lambda_b}^{exp}=(5620.2\pm 1.6)~{\rm MeV}~,
\eea
we deduce:
\bea
M_{\Xi_b}=(5789\pm 16)~{\rm MeV}~,
\eea
which agrees quite well with the data \cite{PDG}:
\bea
M_{\Xi_b}^{exp}=(5792.4\pm 3.0)~{\rm MeV}~.
\eea
Allowing a 1$\sigma$ deviation from the data, we deduce:
\bea
\kappa= 0.738(23)~. 
\label{eq:kchib}
\eea

 \subsection*{\b \boldmath $\Omega_c(css)/\Sigma_c(cqq)$}
 \nin
\begin{figure}[hbt]
 \vspace*{-0.5cm}
\begin{center}
\includegraphics[width=6cm]{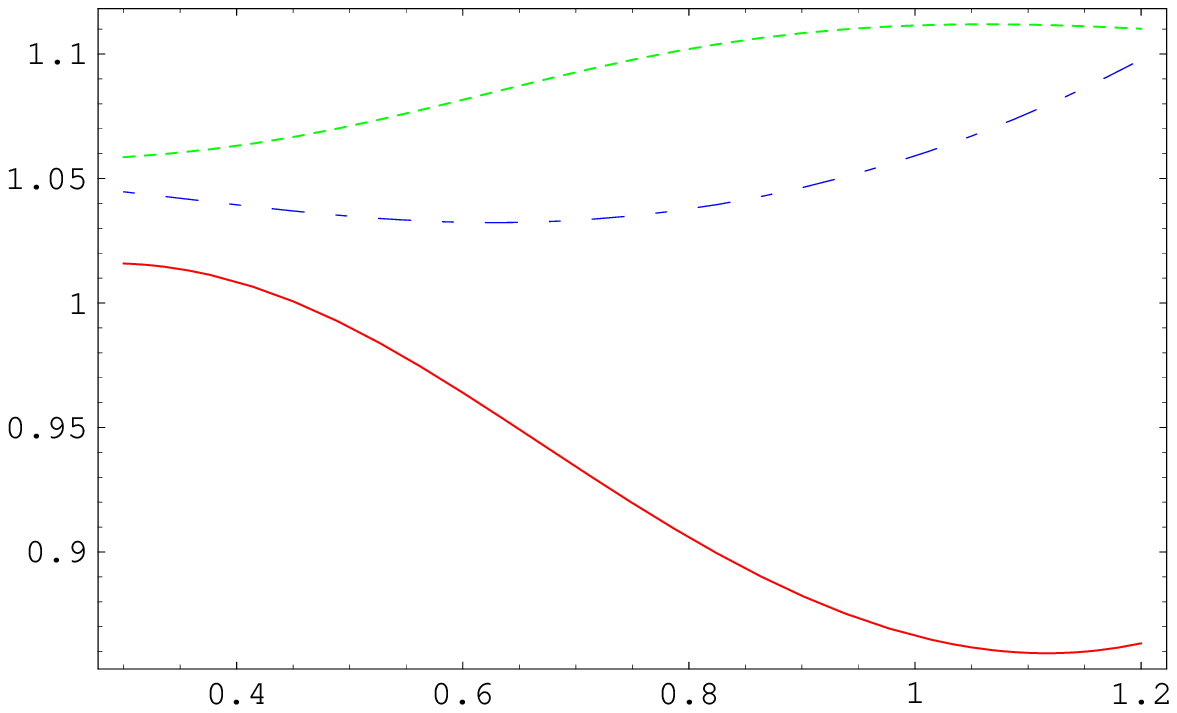}
\includegraphics[width=6cm]{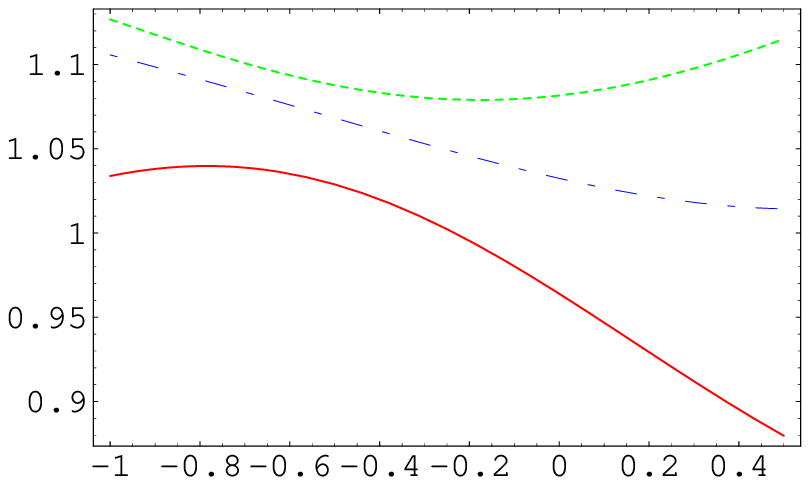}
\includegraphics[width=6cm]{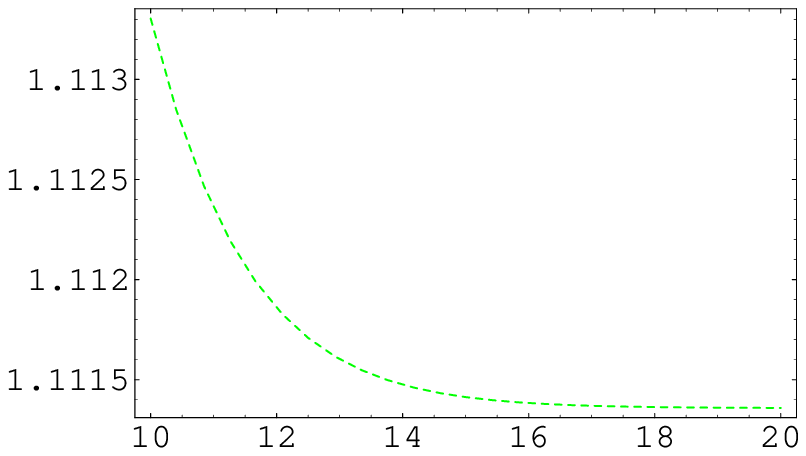}
\vspace*{-0.7cm}
\caption{\footnotesize {\boldmath $\Omega_c/\Sigma_c$}: a) 
$\tau$-behaviour of DR given $b=0$ and $tc=14$ GeV$^{2}$: $r_1^{ds}$ dashed-dotted (blue),  $r_2^{ds}$ dotted (green), $r_{21}^{ds}$ continuous (red);
b) $b$-behaviour of the DR given $\tau=0.6$ GeV$^{-2}$ ($\tau$-stability of $r_1^{ds}$) and $t_c=14$ GeV$^2$; 
c) $t_c$-behaviour of $r_2^{ds}$ given $b=0$ and $\tau=1$ GeV$^{-2}$. }
\label{fig:omegac}
\end{center}
\vspace*{-1.cm}
\end{figure}
\nin
 We do an analysis similar  to the one in the previous section. The result for the $c$-quark is shown in Fig \ref{fig:omegac} . One can notice that the optimal choice of the current is the same as in Eq. (\ref{eq:mixing1}) which we fix to the value $b$=0. One can notice from Fig \ref{fig:omegac}a) and Fig \ref{fig:omegac}b), that  only $r_2^{sd}$ presents simultaneously $\tau$ and $b$ stabilities from which we extract the optimal result.   Using $\kappa=0.78$, the final  result from $r_2^{sd}$ is:
  \beq
 r_{\Omega_c}^{sd}=1.111(1.4)(1.3)(16.4)(0.2)~.
 \eeq
  The errors are due respectively to the values of  $\tau=(1.0\sim 1.2)$ 
GeV$^{-2}$, $m_c$, $m_s$ and the factorization of the four-quark condensate $\rho$. 
The one due to $t_c$ from 10 to 20 GeV$^2$ is negligible. 
 Using this previous result together with the experimental averaged value~\cite{PDG}:
 \beq
 M_{\Sigma_c}^{exp}=(2453.6\pm 0.25)~{\rm MeV}~ , 
 \eeq
 one can deduce: 
  \beq
 M_{\Omega_c}=2726.9(40.5)~{\rm MeV}~ , 
 \eeq
 in good agreement with the data:
 \beq
 M_{\Omega_c}^{exp}=(2697.5\pm 2.6)~{\rm MeV}~.
 \eeq
 Now, we study the influence of $\kappa$ on the mass prediction. Allowing a $1\sigma$ deviation from
 the experimental mass, we deduce the estimate:
 \bea
\kappa= 0.775(15)~. 
\label{eq:komegac}
\eea
 \subsection*{\b \boldmath Final value of \boldmath$\kappa$}
 \nin
Taking the (arithmetic) mean value of $\kappa\equiv\la \bar ss\ra/\la \bar 
dd\ra $ from the different channels $\Xi_{c,b}$ [Eqs. (\ref{eq:kchic}) and (\ref{eq:kchib})] and $\Omega_c$ [Eq. (\ref{eq:komegac})], we deduce:
\beq
\kappa= 0.738(29)~,
\label{eq:fkappa}
\eeq
which we can consider as an improved estimate of this quantity compared with the existing one from
the light mesons \cite{SNmass,SNB} and baryons \cite{JAMI2}:
\beq
\kappa= 0.700(100)~.
\eeq
\section{The mass of the \boldmath$\Omega_b(bss)$}
\vspace*{-0.25cm}
\nin
 \begin{figure}[hbt]
 \vspace*{-0.5cm}
\begin{center}
\includegraphics[width=6cm]{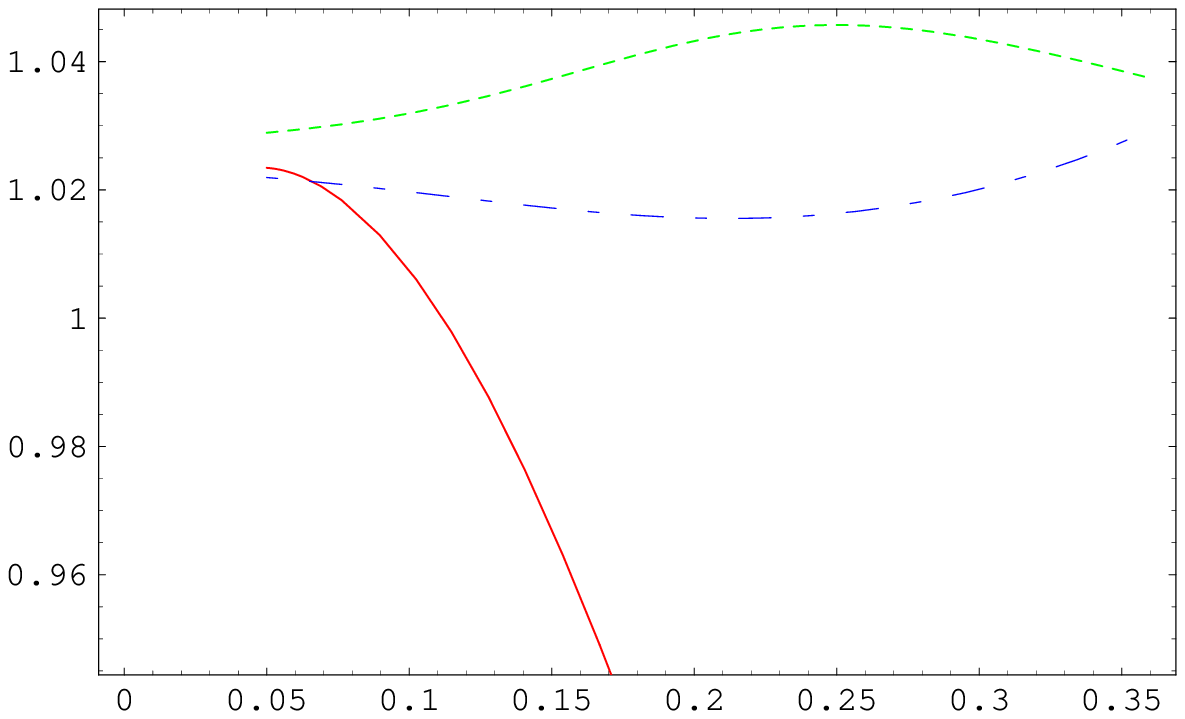}
\includegraphics[width=6cm]{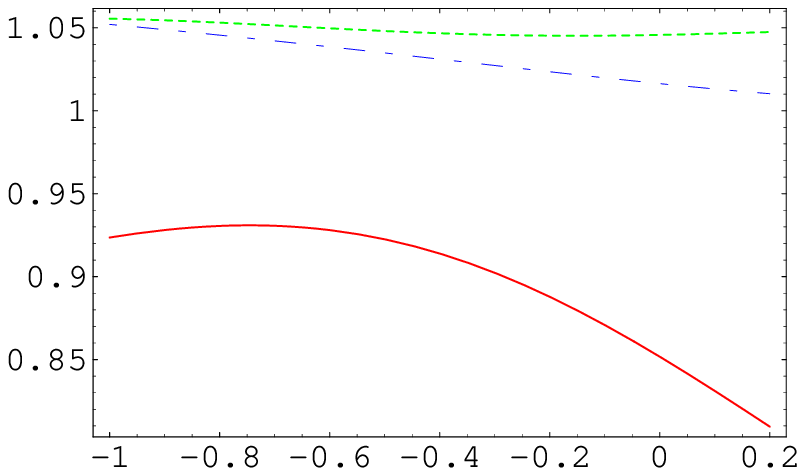}
\includegraphics[width=6cm]{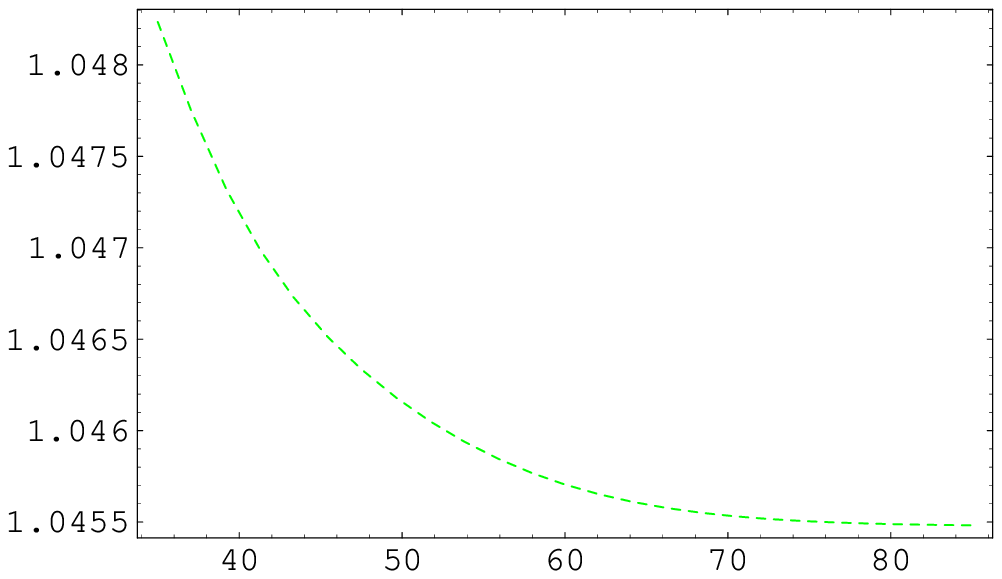}
\vspace*{-0.7cm}
\caption{\footnotesize {\boldmath $\Omega_b/\Sigma_b$}: a) 
$\tau$-behaviour of DR given $b=0$ and $tc=60$ GeV$^{2}$: $r_1^{ds}$ dashed-dotted (blue),  $r_2^{ds}$ dotted (green), $r_{21}^{ds}$ continuous (red);
b) $b$-behaviour of the DR given $\tau=0.25$ GeV$^{-2}$ ($\tau$-stability of $r_1^{ds}$) and $t_c=60$ GeV$^2$; 
c) $t_c$-behaviour of $r_2^{ds}$ given $b=0$ and $\tau=0.25$ GeV$^{-2}$. We have used $\kappa=0.738.$}
\label{fig:omegab}
\end{center}
\vspace*{-1.cm}
\end{figure}
\nin
We repeat the previous analysis  of the $\Omega_c(css)$ in the case of the $b$-quark. The curves present the same qualitative behaviour as in the case of the charm, where, only  $r_{2}^{ds}$ survives the different tests of stabilities. The optimal value is taken at the extremum $\tau=(0.25\pm 0.05)$ GeV$^{-2}$ and in the $t_c$ stability region. Then, we obtain:
 \beq
 r_{\Omega_b}^{sd}=1.0455(20)(22)(41)(13)(37)~.
 \eeq
  The errors are due respectively to the values of  $\tau$, $m_b$, $m_s$ and the factorization of the four-quark condensate $\rho$. 
The last error is due to $\kappa$. The one due to $t_c$ in the $t_c$ stability region is negligible. 
 Using this value together with the experimental averaged value \cite{PDG}:
 \beq
M_{\Sigma_b}^{exp}=5811.2~{\rm MeV},
\eeq
 one can deduce  the result:
 \beq
M_{\Omega_b}=6075.6(37.2)~{\rm MeV}~,
\eeq
which we compile in  Table \ref{tab:mass}.
This result agrees within the errors with the one from the CDF collaboration \cite{CDF}: $6054.4(6.9)~{\rm MeV}$ 
but disagrees by about 2.4 $\sigma$ with the D0 value \cite{ABAZOV}: $M_{\Omega_b}^{D0}=6165.0(13.0)~{\rm MeV}$  given in the same table.

\section{The mass of the \boldmath$\Xi'_{c,b}(Qsq)$}
\begin{figure}[hbt]
 \vspace*{-0.5cm}
\begin{center}
\includegraphics[width=6cm]{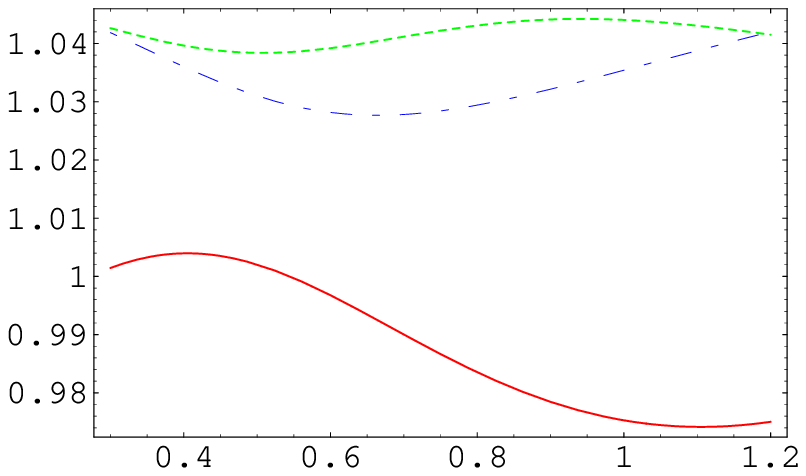}
\includegraphics[width=6cm]{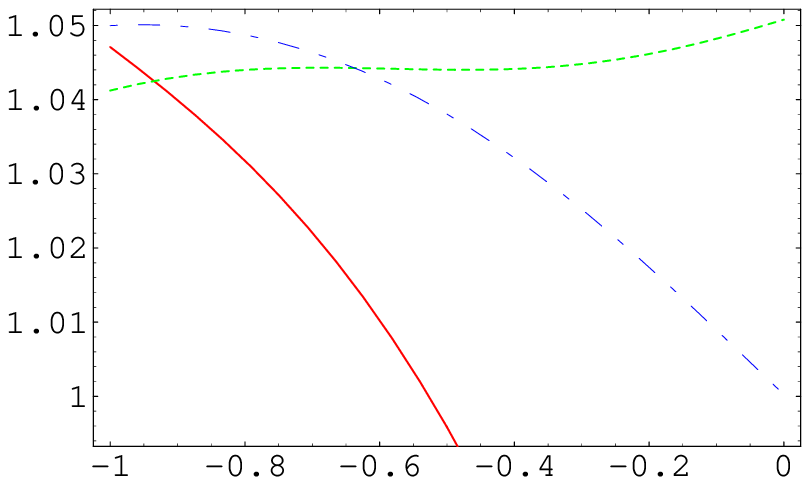}
\includegraphics[width=6cm]{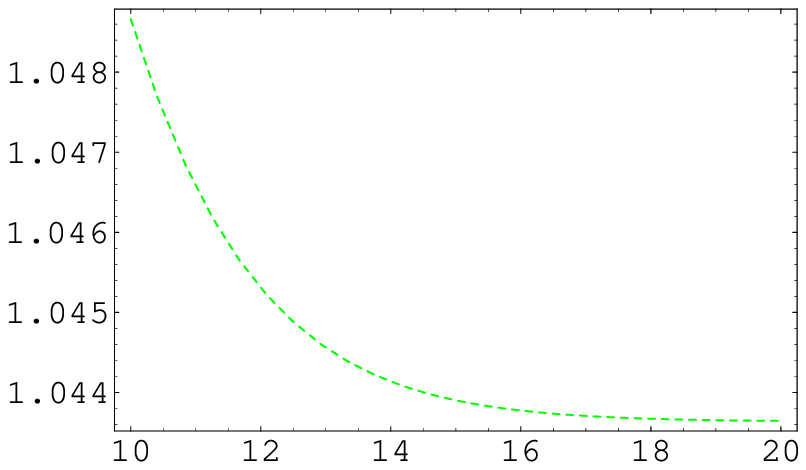}
\vspace*{-0.7cm}
\caption{\footnotesize {\boldmath $\Xi'_{c}/\Sigma_c$}: a) 
$\tau$-behaviour of DR given $b=-0.4$ and $tc=14$ GeV$^{2}$: $r_1^{ds}$ dashed-dotted (blue),  $r_2^{ds}$ dotted (green), $r_{21}^{ds}$ continuous (red);
b) $b$-behaviour of the DR given $\tau=0.9$ GeV$^{-2}$ and $t_c=14$ GeV$^2$; 
c) $t_c$-behaviour of $r_2^{ds}$ given $b=-0.4$ and $\tau=1$ GeV$^{-2}$. We have used $\kappa=0.74.$}
\label{fig:chiprimc}
\end{center}
\vspace*{-1.cm}
\end{figure}

\begin{figure}[hbt]
 \vspace*{-0.5cm}
\begin{center}
\includegraphics[width=6cm]{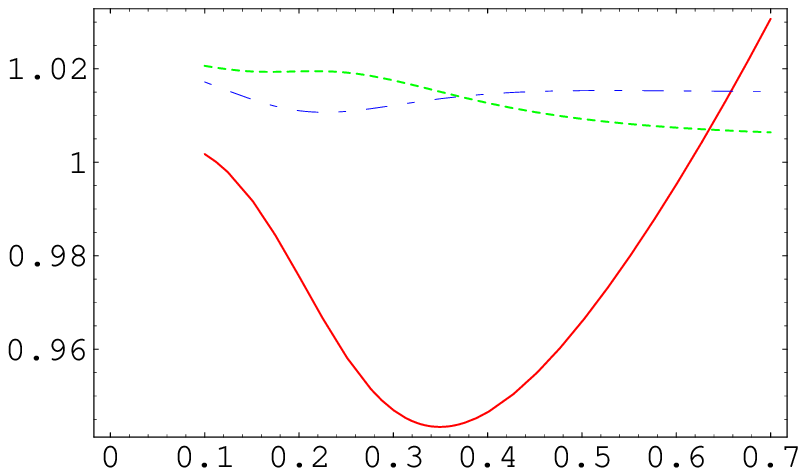}
\includegraphics[width=6cm]{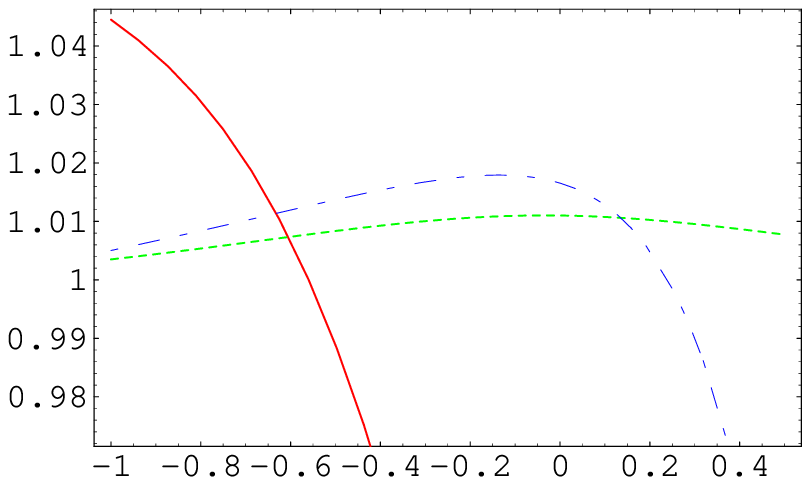}
\includegraphics[width=6cm]{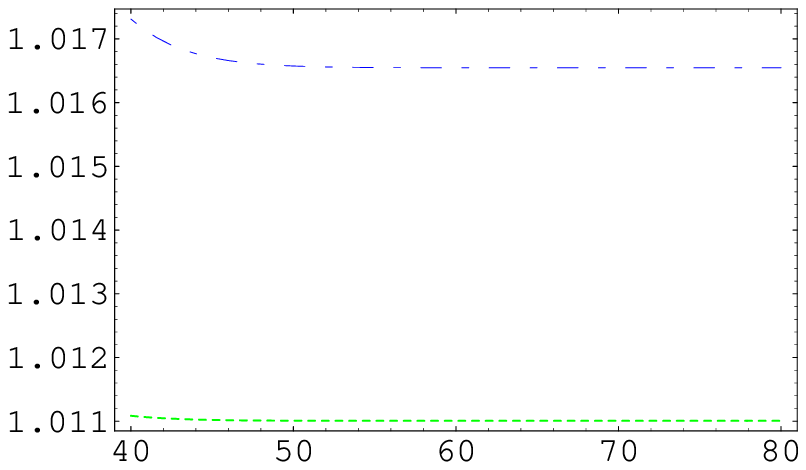}
\vspace*{-0.7cm}
\caption{\footnotesize {\boldmath $\Xi'_{b}/\Sigma_b$}: a) 
$\tau$-behaviour of DR given $b=0$ and $tc=60$ GeV$^{2}$: $r_1^{ds}$ dashed-dotted (blue),  $r_2^{ds}$ dotted (green), $r_{21}^{ds}$ continuous (red);
b) $b$-behaviour of the DR given $\tau=0.5$ GeV$^{-2}$ and $t_c=60$ GeV$^2$; 
c) $t_c$-behaviour of $r_2^{ds}$ given $b=0$ and $\tau=0.5$ GeV$^{-2}$. We have used $\kappa=0.74.$}
\label{fig:chiprimb}
\end{center}
\vspace*{-1.cm}
\end{figure}

\vspace*{-0.25cm}
\nin
We do a similar analysis for the $\Xi'_c$, which we show in Fig. \ref{fig:chiprimc}, where we have only retained the $r_2^{ds}$ which satisfies all stability tests. 
\nin
\vspace*{-0.25cm}
We obtain:
 \beq
 r_{\Xi'_c}^{sd}=1.043(1)(2)(6)(2)(3)(7)~,
 \eeq
  The errors are due respectively to the values of  $\tau=(0.9\pm 0.1)$ 
GeV$^{-2}$, $m_c$, $m_s$, $\rho$, $b=-(0.4\pm 0.2)$, and $\kappa$. Using the experimental value of $M_{\Sigma_c}$, we obtain:
 \beq
M_{\Xi'_c}=2559(25)~{\rm MeV},
\eeq
which we compile in Table \ref{tab:mass}. Our prediction is in good agreement ($1\sigma$) with the data \cite{PDG}:
 \beq
M_{\Xi'_c}^{exp}=2576(3.1)~{\rm MeV}~.
\eeq
A similar analysis is done for $\Xi'_b$, which is sumarized in Fig. \ref{fig:chiprimb}.  One can notice  that here the $b$-stability occurs at 0 in this channel and there is no sharp selection between $r_1^{ds}$ and $r_2^{ds}$.
\nin
We obtain the mean from $r_1^{ds}$ and $r_2^{ds}$:
 \beq
 r_{\Xi'_b}^{sd}=1.014(3.4)(5)(1.7)(0.5)(2)(0.5)(3)~,
 \eeq
 where the sources of the errors are $\tau=(0.5\pm 0.1)$ 
GeV$^{-2}$, $m_b$, $m_s$, $\rho$, $b=(0.\pm 0.2)$, and $\kappa$. The last error comes from the choice of the sum rules. Using the experimental value of $M_{\Sigma_b}$, we predict:
 \beq
M_{\Xi'_b}=5893(42)~{\rm MeV}~,
\eeq
which we compile in Table \ref{tab:mass}.

\section{The masses of the spin 3/2 baryons}
\vspace*{-0.25cm}
\nin
As a preliminary step of the analysis, we check the different results obtained in \cite{BAGAN2}:
\bea
&& M_{\Sigma^*_c}=(2.15\sim 2.92)~{\rm GeV}~,\nnb\\
&& M_{\Sigma^*_b}-M_{\Sigma^*_c}=3.3~{\rm GeV}~,
\eea
and confirm them.
However, like in the spin 1/2 case, we have not tried to improve these (old) results. 
 \subsection*{\b \boldmath $\Xi_c^*(csq)/\Sigma_c^*(cqq)$ }
\begin{figure}[hbt]
   \vspace*{-0.5cm}
\begin{center}
\includegraphics[width=6cm]{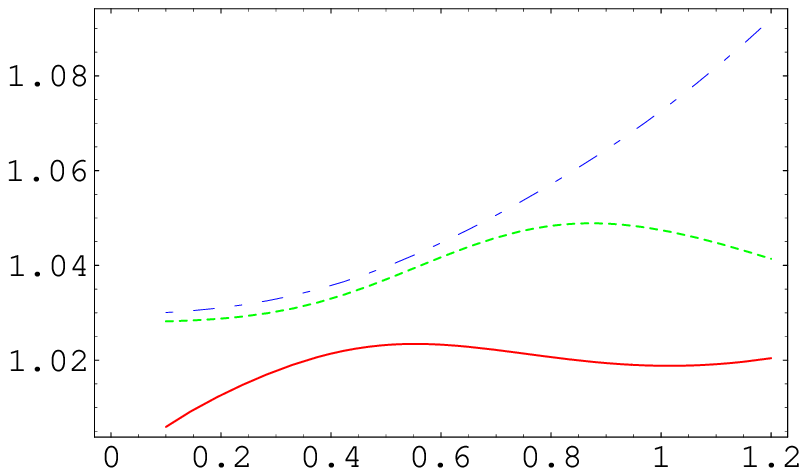}
\includegraphics[width=6cm]{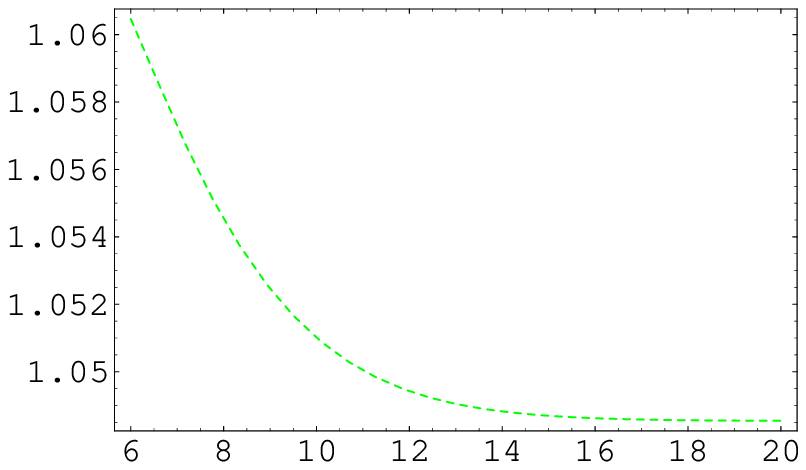}
\vspace*{-0.7cm}
\caption{\footnotesize {\boldmath $\Xi^*_c/\Sigma^*_c$}:  a) $\tau$-behaviour of the double ratio of sum rules (DR) by giving $t_c=14$ GeV$^2$ : $r_1^{ds}$ dashed-dotted (blue),  $r_2^{sd}$ dotted (green) and  $r_{21}^{sd}$ continuous (red); 
c) $t_c$-behaviour of  $r_2^{ds}$ for a given optimal $\tau=0.9$ GeV$^{-2}$. }
\label{fig:chic*}
\end{center}
\end{figure}

 \nin
We repeat the previous DR analysis for the case of the $\Xi_c^*$. We show in Fig. \ref{fig:chic*}a)  the $\tau$-behaviour of the mass predictions for  $t_c$ = 14 GeV$^2$. From this analysis, we do not retain $r_{21}^{sd}$ which differs completely from $r_1^{ds}$ and $r_2^{ds}$, while we do not consider $r_1^{ds}$ which is  $\tau$-instable. We show in Fig. \ref{fig:chic*}b) the $t_c$-behaviour of $r_2^{ds}$ given $\tau$=0.9 GeV$^{-2}$. 
\\
\nin
We deduce the optimal value:
\bea
r_{\Xi^*_c}^{sd}= 1.049(1)(10)(4)(4)(17.5)(1)~.
\eea
The errors are due respectively to the values of $\tau=(0.9\pm 0.1)$ GeV$^{-2}$, $m_c$, $m_s$, $\rho$ (factorization of the four-quark condensate) and  $\kappa\equiv \la \bar ss\ra/\la \bar dd\ra=0.74\pm 0.03$. The ones due to some other parameters are negligible.
Using the data \cite{PDG}:
\bea
M_{\Sigma^*_c}^{exp}=(2517.97\pm 1.17)~{\rm MeV}~,
\label{data_sigma*}
\eea
and adding the different errors quadratically, we deduce the results in Table \ref{tab:mass}.
\subsection*{\b  \boldmath $\Xi_b^*(bsq)/\Sigma_b^*(bqq)$}
\nin
We extend the analysis to the case of the bottom quark.  The curves are qualitatively similar to the charm case. 
We deduce:
\bea
r_{\Xi^*_b}^{sd}= 1.022(2)(2)(0.5)(1)(2)~.
\eea
The sources of the errors are the same as for the $\Xi^*_c$, where here $\tau=(0.25 \pm 0.05)$ GeV$^{-2}$ and $m_c$ replaced by $m_b$.The ones due to some other parameters are negligible.
Using the averaged data \cite{PDG}:
\bea
M_{\Sigma^*_b}^{exp}=(5832.7\pm 6.5)~{\rm MeV}~,
\label{eq:sigma*b}
\eea
 and adding the different errors quadratically, we deduce:
 \bea
M_{\Xi^*_b}=(5961\pm 21)~{\rm MeV}~,
\label{eq:chib*}
\eea
 which we report in Table \ref{tab:mass}.  
\subsection*{\b  \boldmath $\Omega^*_c(css)/\Sigma^*_c(cqq)$}
 \nin
 \begin{figure}[hbt]
  \vspace*{-0.5cm}
\begin{center}
\includegraphics[width=6cm]{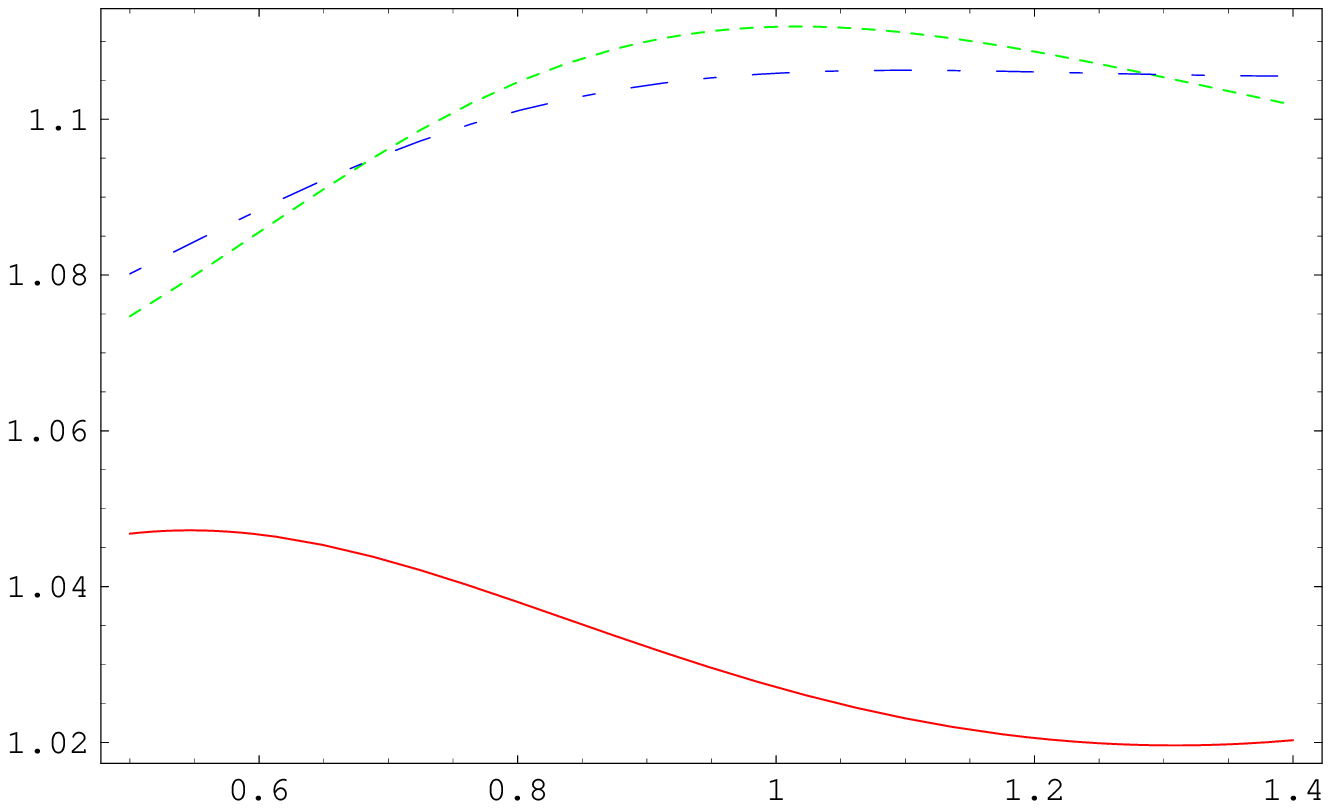}
\includegraphics[width=6cm]{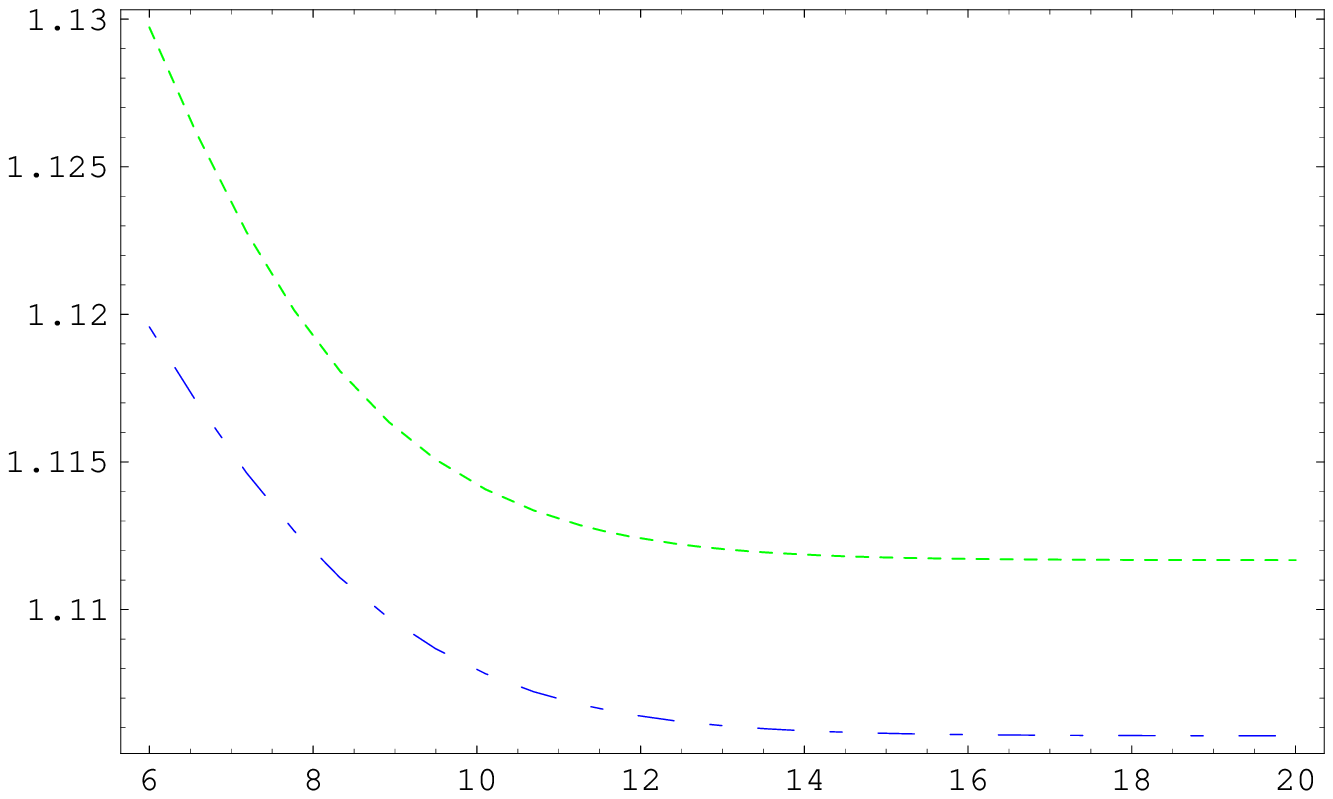}
\vspace*{-0.7cm}
\caption{\footnotesize {\boldmath $\Omega^*_c/\Sigma^*_c$}:  a) $\tau$-behaviour of the double ratio of sum rules (DR) by giving $t_c=14$ GeV$^2$ : $r_1^{ds}$ dashed-dotted (blue),  $r_2^{sd}$ dotted (green) and $r_{21}^{sd}$ continuous (red); 
b) $t_c$-behaviour of the DR  $r_1^{ds}$ and $r_2^{ds}$ giving $\tau=1$ GeV$^{-2}$. }
\label{fig:omegac*}
\end{center}
\vspace*{-1.cm}
\end{figure}
\nin
We pursue the analysis to the case of the $\Omega^*_c(css)$. We show  the $\tau$-behaviour of the different DR
 in Fig. \ref{fig:omegac*}a). From this figure, we shall not retain $r_{21}^{ds}$ which differs completely from $r_1^{ds}$ and $r_2^{ds}$.  Given the optimal value of $\tau=1$ GeV$^{-2}$, we show the $t_c$-behaviour of the DR $r_1^{ds}$ and $r_2^{ds}$ in Fig. \ref{fig:omegac*}b). The final result is the mean from $r_1^{ds}$ and $r_2^{ds}$:
\bea
r_{\Omega^*_c}^{sd}= 1.109(3)(10)(10)(4)(0.5)(8)~.
\eea
The 1st error is due to the choice of $r_i^{sd} $. The other ones are due to $\tau=(1.0\pm 0.2)$ GeV$^{-2}$, $m_c$, $m_s$,  $\rho$ and $\kappa$. The other QCD parameters give negligible errors. Using the averaged data in Eq. (\ref{eq:sigma*b}), and adding the different errors quadratically, one can deduce the result in Table~\ref{tab:mass}.  
\subsection*{\b  \boldmath $\Omega^*_b(bss)/\Sigma^*_b(bqq)$}
 \nin
We repeat the previous analysis in the $b$-channel. The curves are qualitatively analogue to the ones of the charm. We shall not consider $r_{21}^{sd}$ because of its incompatibility with the other ones. From the mean of $r_1^{ds}$ and  $r_2^{ds}$, we deduce:
\bea
r_{\Omega^*_b}^{sd}= 1.040(4)(2)(4.6)(0.2)(6)~,
\eea
where the sources of the errors are the same as for $\Omega^*_c$, where $\tau=(0.30\pm 0.05)$ GeV$^{-2}$ here and $m_c$ replaced by $m_b$. Using the averaged data in Eq. (\ref{eq:sigma*b}), and adding the different errors quadratically, one can deduce:
\bea
M_{\Omega^*_b}=(6066\pm 49)~{\rm MeV}~,
\label{eq:omegab*}
\eea
 which we report in Table \ref{tab:mass}.  
{\scriptsize
\begin{table}[hbt]
\setlength{\tabcolsep}{0.35pc}
 \caption{\scriptsize    QSSR predictions of the strange heavy baryon masses in units of MeV from the double ratio (DR) of sum rules with the QCD input parameters in Table \ref{tab:param} and using as input the observed masses of the associated non-strange heavy baryons. We have used $\kappa\equiv \la \bar ss\ra/\la \bar dd\ra=0.74\pm 0.03$ fixed from the experimental $\Xi_{c,b}$ and $\Omega_c$ masses. }
 {\small
\begin{tabular}{lllll}
&\\
\hline
Baryons&$I$&$r_{B^*_Q}^{sd}$&Mass&Data   \\
\hline
\\
{ $J^P={1\over 2}^+$}&&&&\\
&\\
$\Xi_c(cqs)$&${1\over 2}$&&input&$2467.9\pm 0.4$ [PDG]\\
$\Omega_c(css)$&${0}$&&input&$2697.5\pm 2.6$ [PDG]\\
$\Xi_b(bqs)$&${1\over 2}$&&input&$5792.4\pm 3.0$ [PDG]\\
$\Xi'_c(cqs)$&${1\over 2}$&1.043(10)&2559(25)&$2575.7\pm 3.1$ [PDG]\\
$\Xi'_b(bqs)$&${1\over 2}$&1.014(7)&5893(42)&$-$\\
$\Omega_b(bss)$&${0}$&1.0455(64)&6076(37)&$6165.0\pm 13$ [D0]\\
&&&&$6054.4\pm 6.9$ [CDF]\\

%
{ $J^P={3\over 2}^+$}&&&& \\
&\\
$\Xi^*_c(cqs)$&${1\over 2}$&1.049(8)&2641(21)&$2646.1\pm 1.3$ [PDG]\\
$\Omega^*_c(css)$&0&1.109(17)&2792(38)&$2768.3\pm 3.0$ [PDG]\\
$\Xi^*_b(bqs)$&${1\over 2}$&1.024(8)&5961(21)&$-$ \\
$\Omega^*_b(bss)$&0&1.040(9)&6066(49)&$-$\\

\hline
\end{tabular}
\vspace*{-0.5cm}
}
\label{tab:mass}
\end{table}
}
 \vspace*{-0.25cm}
{\scriptsize
\begin{table}[hbt]
\setlength{\tabcolsep}{1.8pc}
 \caption{\scriptsize    QSSR predictions of the strange heavy baryon hyperfine splittings in units of MeV from the double ratio (DR) of sum rules with the QCD input parameters in Table \ref{tab:param} and using as input the predicted values in Table \ref{tab:mass}. We have added the errors quadratically.  }
 {\small
\begin{tabular}{ll}
&\\
\hline
Hyperfine Splittings& Data   \\
\hline
$M_{\Xi^*_c}-M_{\Xi_c}$=173(21)&$179(1)$\\
$M_{\Xi^*_c}-M_{\Xi'_c}$=82(33)&$70(3)$\\
$M_{\Omega^*_c}-M_{\Omega_c}$=95(38)&70(3)\\
%
%
$M_{\Xi^*_b}-M_{\Xi_b}$=169(21)&$-$ \\
$M_{\Xi^*_b}-M_{\Xi'_b}$=68(47)&$-$ \\
$M_{\Omega^*_b}-M_{\Omega_b}$=-10(61)&$M_{\Sigma^*_b}-M_{\Sigma_b}=22$\\
%
\hline
\end{tabular}
}
\vspace*{-0.5cm}
\label{tab:hyperfine}
\end{table}
}
\nin
 \section{ Hyperfine mass-spilltings}
 \vspace*{-0.25cm}
 \nin
Combining the results for spin 1/2 and spin 3/2 given in Table \ref{tab:mass}, we deduce in Table \ref{tab:hyperfine} the values of the hyperfine mass-spilittings. From our analysis, one expects that the $\Omega^*_Q$ can only decay electromagnetically to $\Omega_Q+\gamma$ due to phase space, while the $\Xi^*_Q$ can, in addition, decay hadronically to $\Xi_Q+\pi$. On one hand, our result for the unobserved mass-differences $M_{\Omega^*_b}-M_{\Omega_b},~M_{\Xi^*_b}-M_{\Xi'_b}$ agree within the errors with the ones from quark models \cite{RICHARD} and $1/N_c$ expansion \cite{JENKINS} and seems to behave like $1/m_b$ despite the large errors.  On the other hand, we predict with a better precision $M_{\Xi^*_b}-M_{\Xi_b}\simeq M_{\Xi^*_c}-M_{\Xi_c}$. A future precise measurement of the $\Xi'_b,~\Xi^*_b$ and $\Omega^*_b$ will shed light on the quark mass behaviour of these mass-differences, which we also plan to study in a future work.
 \section{ Summary and Conclusions}
 \vspace*{-0.25cm}
 \nin
 We have directly extracted (for the first time) the heavy baryons (charmed $C=1$ and bottom $B=-1$) mass-splittings due to $SU(3)$ breaking   using double ratios (DR) of QCD spectral sum rules(QSSR), which are less sensitive to the exact 
value and the definition of the heavy quark mass, to the perturbative radiative
corrections and to the QCD continuum contributions than the simple ratios commonly used in the current literature for determining the 
heavy baryon masses: \\
\b Remarking that the leading term controlling the mass-splittings is, in most of the cases, the ratio $\kappa\equiv \la \bar ss\ra/\la \bar dd\ra$ of the  condensate rather than the running mass $\bar{m}_s$, we use as input the observed masses of the $\Xi_{c,b}$ and $\Omega_c$, for extracting  $\kappa$. We obtain the mean value from Eqs. (\ref{eq:kchic}), (\ref{eq:kchib}) and  (\ref{eq:komegac}):
\beq
\kappa= 0.738(29)~ {\rm [Eq.~ (\ref{eq:fkappa})]}~,
\eeq
which we can consider as an improved estimate of this quantity compared with the existing one $\kappa= 0.7\pm 0.1$ compiled in Table  \ref{tab:param} from
the light mesons \cite{SNmass,SNB} and baryons \cite{JAMI2}.\\
\b Using this value of $\kappa$, we give predictions of the $\Xi'_{c,b}, ~\Omega_b$ and spin 3/2 baryons masses which  are summarized in Table \ref{tab:mass}. These predictions are in good agreement with the experimental masses and can be considered as improvements of
existing QSSR results based on the simple ratio of moments \cite{MARINA,SR}.\\
\b Our result for the $\Omega_b$ favours the one observed by CDF \cite{CDF} but disagrees within 2.4$\sigma$ with the one from D0 \cite{ABAZOV}. \\
\b Most of predictions agree with the  ones from different approaches (quark models \cite{RICHARD},  lattice calculations \cite{LATT}
and large $N_c$ \cite{JENKINS}). Our predictions for the not yet observed states $\Xi'_b,~\Xi^*_b$ and $\Omega^*_b$ given in Table \ref{tab:mass} can serve in a near future as a test of the QSSR approach.  \\
\b We show in Table \ref{tab:hyperfine} our predictions for the hyperfine splittings. Our results agree with the observed values, while the ones for not yet observed states can serve as a test of the QSSR approach. From our results, we expect that the $\Omega^*_Q$ can only decay electromagnetically to $\Omega_Q+\gamma$ due to the available phase space, while the $\Xi^*_Q$ can, in addition, decay hadronically to $\Xi_Q+\pi$. \\
\b One can also notice that, if we have used a $SU(3)$ symmetric quark condensates $\la \bar ss\ra\simeq \la \bar dd\ra$, the predictions would be systematically lower by about $(70\sim 100)$ MeV than the predictions given in Table \ref{tab:mass}.  In this case, the agreement with the observed masses in different channels cannot be achieved.  It would be interesting to understand analogous effects  of $\kappa$ using some other approaches. 
\vspace*{-0.35cm}
\section*{Acknowledgements} 
\vspace*{-0.25cm}
\nin
R.M.A and M.N. acknowledge the  financial support from FAPESP and CNPq (Brazil). M.N. also thanks S. Narison for the hospitality at the Laboratoire de Physique Th\'eorique et Astroparticules of Montpellier. S.N. has been partially  supported by the CNRS-IN2P3 within the project ``Hadron Physics in QCD" and within the French-China Particle Physics Laboratory (FCPPL). We thank Jean-Marc Richard for some discussions.

\vspace*{-0.25cm}


\end{document}